\PassOptionsToPackage{bookmarks=false}{hyperref}
\documentclass[conference]{IEEEtran}
\IEEEoverridecommandlockouts
\usepackage{hyperref}
\hypersetup{draft}
\usepackage{cite}
\usepackage{algorithmic}
\usepackage{graphicx,ragged2e}
\usepackage{graphics}
\usepackage{comment}
\usepackage{graphicx}
\usepackage{hyperref}
\usepackage{xcolor}
\usepackage{subcaption}
\usepackage{amsmath,amssymb}
\usepackage{float}
\usepackage{caption}

\DeclareCaptionType{equ}[][]

\usepackage{abraces}
\def\BibTeX{{\rm B\kern-.05em{\sc i\kern-.025em b}\kern-.08em
		T\kern-.1667em\lower.7ex\hbox{E}\kern-.125emX}}

\newtheorem{remark}{Remark}[section]

\begin{document}

\title{Formulating the Restoration of Distribution Networks as a Multiple Traveling Salesman Problem}
\author{Ran Wei, 
        ~\IEEEmembership{Graduate Student Member,~IEEE},
        Arindam~K.~Das, ~\IEEEmembership{Member,~IEEE},
        Daniel~S.~Kirschen, ~\IEEEmembership{Life Fellow,~IEEE}, \\
        and~Payman~Arabshahi, ~\IEEEmembership{Senior~Member,~IEEE}
\thanks{This research was supported by the National Science Foundation under Grant No. 2139837: ``Optimal restoration of electricity distribution networks under rolling time windows and prediction of restoration time''.}
\thanks{R. Wei, P. Arabshahi and D. S. Kirschen are with the Department of Electrical and Computer Engineering, University of Washington, Seattle, WA, 98195-2500 USA. (Email: \href{mailto:rawe1722@uw.edu }{rawe1722@uw.edu })}
\thanks{A. K. Das is with the Department of Computer Science and Electrical Engineering, Eastern Washington University, Cheney, WA, 99004-2493, USA.}
}
\maketitle
%

\begin{abstract}
%
Severe weather events can cause extensive damage to electrical distribution networks, requiring a multi-day restoration effort. 
Optimizing the dispatch of repair crews minimizes the severe socio-economic consequences of such events.
Considering both repair times and travel times, we use graphical manipulations to transform this multiple crew scheduling problem into a type of traveling salesman problem(TSP). Specifically, we demonstrate that the restoration problem bears major resemblance to an instance of a \emph{cost constrained reward maximizing} $m$TSP (multiple TSP) \emph{on node and edge weighted} (doubly weighted) graphs (a variant we dub the CCRM-$m$TSP-DW), where the objective is to maximize the aggregate reward earned during the upcoming restoration window, provided no crew violates its time budget and electrical continuity constraints are met. Despite the rich history of research on the TSP and its variants, this CCRM-$m$TSP-DW variant has not been studied before, although its closest cousin happens to be the ``Selective TSP'' (S-TSP). This reinterpretation of the restoration problem not only opens up the possibility of drawing on existing solution methods developed for the TSP and its variants, it also adds a new chapter in the annals of research on ``TSP-like'' problems. In this paper, we propose a ``TSP-like'' mixed integer linear programming (MILP) model for solving the restoration problem and validate it on the IEEE PES 123-node test feeder network. 
\end{abstract}
%
\begin{IEEEkeywords}
Power distribution network, resilience, system restoration, optimal scheduling, rolling horizon, traveling salesman problem (TSP) variant.
\end{IEEEkeywords}
\IEEEpeerreviewmaketitle

\section{Introduction}
Typhoons, ice storms and other natural disasters regularly cause severe damage to distributed power networks. \cite{panteli2015influence}. Because repairing this damage and restoring power often requires a considerable amount of time, the socio-economic consequences of such events are severe \cite{storm2023, storm2023news, campbell2012weather, mohamed2019proactive, hines2009large}. Optimizing the post-disaster recovery is thus critically important. Several factors make this task particularly complex: (i) the large number of components that must be repaired, (ii) the limited number of repair crews, (iii) the time that these crews need to move from one repair site to another, (iv) the constraints on the amount of time that crews are allowed to work, (v) the impact of the topology of the distribution network on the scheduling of the repairs, and (vi) the need to prioritize the re-energization of some loads (e.g., hospitals) \cite{kinn2014extent}.

This paper makes the following contributions:

\begin{itemize}
    \item It frames the optimal repair and restoration of the distribution network as a \emph{new version of the traveling salesman problem} (TSP) \emph{on a doubly weighted graph}, specifically, an instance of a ``cost constrained reward maximizing multiple traveling salesman problem ($m$TSP) on doubly weighted graphs (CCRM-$m$TSP-DW)''. This framing supports a consistent and efficient consideration of all the constraints mentioned above. 

    \item The restoration problem is modeled as a Mixed Integer Linear Program (MILP) which incorporates the required variables in a compact manner. In addition, a simple customization of the cost function through incorporation of a penalty parameter (which could be time-dependent) introduces an element of \emph{fairness} in the repair schedule so that low-priority repairs are not automatically relegated to the back end of the restoration process.

    \item Interpreting the restoration problem as a TSP-like instance is beneficial since it opens up a rich repertoire of research and solution tools (e.g., heuristics or state-of-the-art machine learning methods) for possible adoption.  
    
    \item Our approach is compatible with a rolling optimization framework \cite{rTanetal2019}, which supports the gradual arrival of information about the location and anticipation duration of the repairs. It also supports quick re-optimization in the event of unexpected repair time overruns.
  
\end{itemize}


\section{Literature Review}

Tan et al. \cite{rTanetal2019} established that scheduling the repairs in a distribution network is an NP-hard problem, which implies that obtaining an optimal solution in polynomial time is inherently challenging. Many researchers have since sought to simplify the problem in various ways to achieve either an optimal or near-optimal solution in an acceptable amount of time.

The ``two-stage'' strategy (divide-and-conquer approach) \cite{shi2023preventive} decomposes the repair problem into two sub-optimization models, each with its own objective function. These sub-models are defined as the ``Repair Order'' and the ``Repair Task Assignment'' problems \cite{lu2022distribution}. Fu et al. \cite{fu2021real} divide the distribution network into multiple subsystems and compute optimal solutions in parallel. Sun et al. \cite{sun2022sequential} divide the cyber and physical layers. Regardless of how the two stages are defined, the fact that each sub-problem has its objective function and constraints allows each stage to reach an optimal solution within an acceptable time, ultimately reducing the overall solution time for the entire problem.

Another widely adopted optimization approach is the ``multi-time-step'' approach, where the completions of repair tasks are associated with specific time intervals. For example, \cite{fu2021real, wang2021multi} and \cite{zhang2020sequential} link the completion status of each task to a precise time scale, down to a minute. In \cite{qiu2023hierarchical}, time is divided into minute-long intervals, and each minute is treated as a small, decentralized, partially observable Markov decision process. At the beginning of each minute, an action is taken, making the multi-time-step approach beneficial for subsequent system scheduling problems and handling unexpected situations. When the scope of the repairs is large and thus requires a long restoration time \cite{storm2023, storm2023news}, the time-granularity of the multi-time-step approaches can become a computational bottleneck. 

In the ``event-driven'' approach \cite{sun2022sequential,pang2023dynamic} a power flow calculation is performed after a branch has been repaired. However, based on conversations with industry experts, we concluded that such computations are required only if the repairs modify the configuration of the network.

The authors of \cite{shi2023preventive,qiu2023hierarchical} aim to minimize load shedding costs while the authors \cite{lu2022distribution} focus on the economic losses caused by outages. The second stage of \cite{sun2022sequential} evaluates load loss, while the first stage in both \cite{zhang2020sequential} and \cite{sun2022sequential} aims to minimize load curtailment. Most researchers opt to maximize the total restored load across the entire distribution network, as seen in works such as \cite{fu2021real, zhang2023outage, wang2021multi, pang2023dynamic, guo2022multi}. However, the importance of each fault location varies and is not always proportional to the restored load. For example,  facilities such as police and fire stations require a prompt restoration but have a relatively small load demand. Therefore, an optimization based solely on restored load, load loss, or economic factors may not fully capture the importance of specific locations. 


The computational complexity of this repair scheduling problem is a function not only of the size of the distribution network but also of the number of repairs that must be completed. Since this scheduling has to be performed in real-time, and may have to be performed repeatedly as new information becomes available, an optimization model capable of optimally scheduling a large number of repairs quickly is required. Several features distinguish this work from previous research. First, we do not make the assumption that \emph{all repairs need to be completed within the shortest possible time}. Instead, given a restoration window of arbitrary length, we require a scheduling policy which maximizes the \emph{value accrual} due to the resources expended (crew time), where `value' can be interpreted as simply the number of customers restored or, more generally, some utility/reward function of the customers restored. Second, a finite length restoration window automatically imposes the constraint that it may not be possible to repair all damages  within the window. It is therefore necessary that repairs be picked judiciously to avoid pockets of repaired sections with no ``energization path''. In other words, it should be possible to energize the repaired nodes from the feeder at the end of the restoration window. Accordingly, we include \emph{electrical continuity constraints} in our model. Third, we do not make any assumption regarding the scale of damage. Our model is equally applicable to a fully damaged network as well as a partially damaged network (unlike, for example \cite{zhang2020sequential, pang2023dynamic, lu2022distribution}), the latter requiring additional considerations regarding the electrical continuity constraints discussed above. Finally, since availability of repair crews and their time budgets play a significant role in the restoration timeline, we integrate these \emph{manpower-related} constraints for a comprehensive modeling of the scheduling framework.

\section{Problem Definition}
\label{sec:probDefn}

\begin{figure}[htbp]
  \centering
	\includegraphics[width=\linewidth]{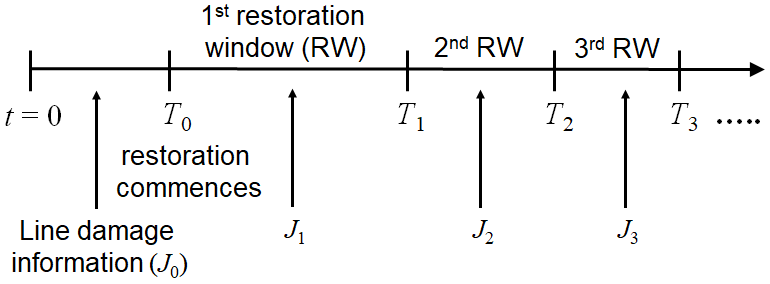}
	\caption{Timeline of the restoration process.} 
\label{fig:restorationTimeline}
\end{figure}

Personnel in charge of optimizing the restoration of the distribution network do not have complete and accurate information about the repairs that must be performed as soon as a storm has passed. Instead, this information arrives gradually. Similarly, not all the repair crews are immediately available. To speed up the restoration process, they must optimally dispatch the available crews based on the information available and periodically re-optimize using updated information.
 Figure~\ref{fig:restorationTimeline} illustrates this rolling horizon restoration framework. Let $t=0$ denote the  onset time of the disaster. Repair and restoration work commences at time $T_0$, after some damage information has been collected and initial repair crews have been mobilized. The time between $T_k$ and $T_{k-1}$ denotes the $k^{th}$ restoration window or planning horizon.  The set of all possible repair jobs during the $k^{th}$ window, $L_k^D$, includes previously known but unscheduled or unfinished jobs from prior windows, as well as new jobs learnt during the $(k-1)^{th}$ window. Dropping the `window subscript' to keep the notations simple, $L^D$ denotes the complete set of repair jobs for any restoration window. Associated with each damaged line is a reward, $r_l$, which we assume is known and determined by the utility company. In this paper, we will use the terms `edge', `line', and `job' interchangeably; `edge' in the context of a graph, `line' in the context of a distribution network, and `job' in the context of a TSP/repair schedule. 

Let $m$ denote the number of repair crews available to work during the $k^{th}$ window. The planning objective for the $k^{th}$ window is to determine which jobs from the set $L^D$ these $m$ crews should perform, such that the total reward that can be earned at the end of the current window is maximized. The cost of any schedule is the sum of the repair times of the jobs on the schedule and the travel times necessary for moving from one job site to the next one. We assume that the travel times between any pair of jobs in $L^D$ are symmetric and correspond to the shortest path between these job sites, as derived from a separate transportation map. Further, we assume that any job site is reachable from any other job site, failing which a job site which is unreachable, possibly due to loss of transportation infrastructure, should be removed from the set $L^D$. Obviously, the cost (length) of any schedule must not exceed the duration of the $k^{th}$ window, $T_k - T_{k-1}$. \emph{Any line, $l$, repaired during the $k^{th}$ window earns the reward $r_l$ only if it can be energized (at the end of the current window) from the previously energized portion of the distribution network. This electrical continuity constraint depends on whether a fully repaired path exists from the energized portion of the distribution network at $T_{k}$ to $l$}. Mathematically, the objective is to construct $m$ repair schedules in order to:
\[ \text{maximize} \ \sum_{l \in L^D} r_l \, I\left(\mathcal{P}(l,S_{k-1})\right), \]
where $S_{k-1}$ denotes the energized portion of the distribution network at time $T_{k-1}$ (the `source'). $I\left(\mathcal{P}(l,S_{k-1})\right)$ is an indicator function which is equal to $1$ if a completely repaired path exists from $l$ to $S_{k-1}$ at time $T_k$ and is equal to $0$ otherwise. Henceforth, we will simply use `path' to mean a `completely repaired path' to the source. If  $r_l=1, \, \forall l$, the objective function reduces to maximizing the number of repairs which can earn an immediate reward at the end of the current restoration window.
\section{Formulation as an $m$TSP}
\label{sec:probFormln}
\begin{figure*}[htb]
  \centering
	\includegraphics[width=0.9\textwidth]{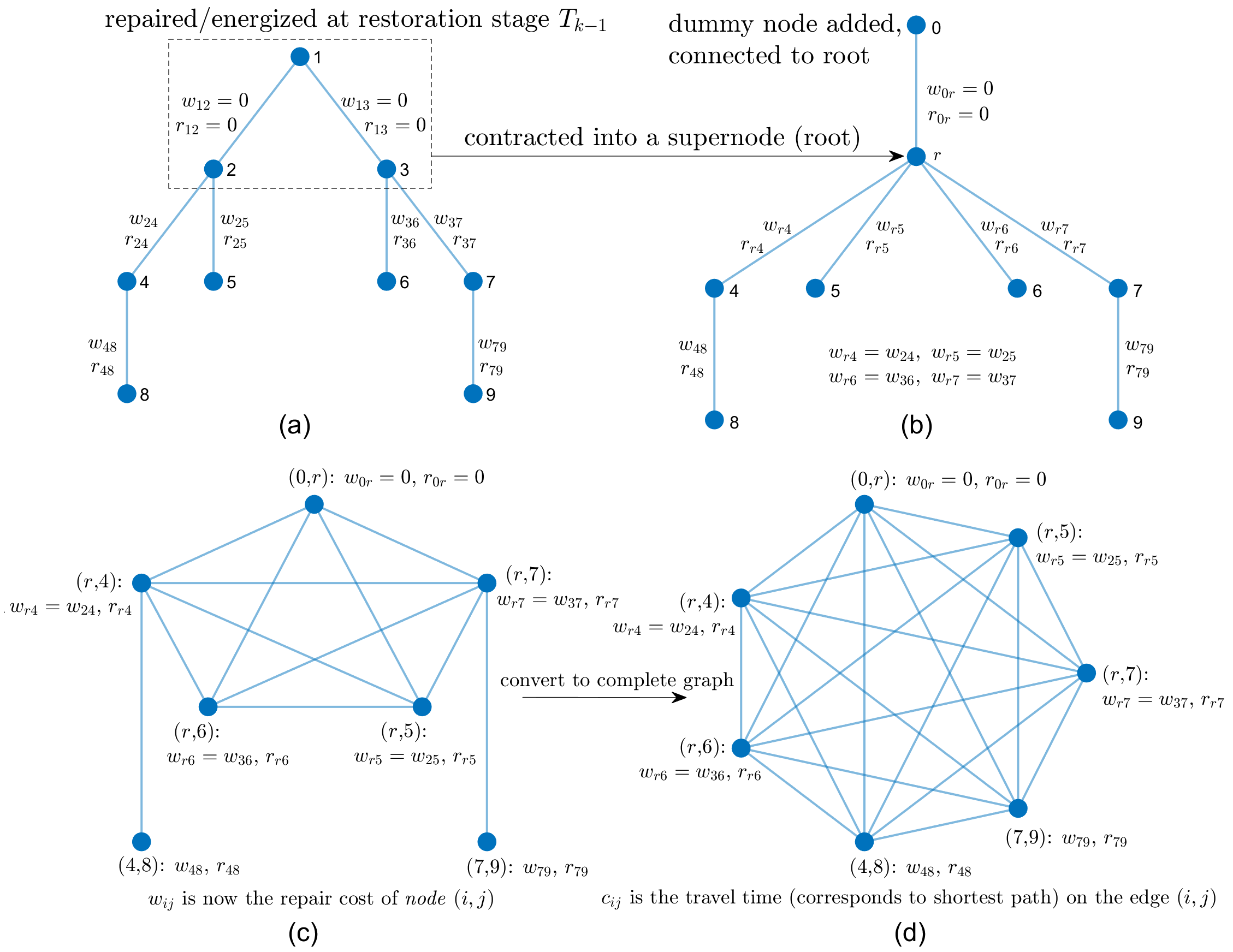}
	\caption{(a) Graph of a radial distribution network, $G = (\mathcal{N},\mathcal{E})$. (b) Modified version of graph $G$, which we denote by $G_m$. (c) Line graph of $G_m$, which we denote by $L(G_m)$. In this figure, we adopt a two-index notation for the node labels in the line graph simply for clarity. (d) The graph $L(G_m)$ converted to a complete symmetric directed graph, which we denote by $L_K(G_m)$. We refer to the node $(0,r)$ as the \emph{root} of $L_K(G_m)$.}
\label{fig:Fig1_2}
\end{figure*}
Consider the radial electrical distribution network, $G$, shown in Figure~\ref{fig:Fig1_2}(a). The dotted portion of the network is assumed to have been repaired and energized at restoration horizon $T_{k-1}$. The other edges in the network constitute the set $L^D$ of job sites. Let $r_{ij}$ denote the reward associated with line $(i,j)$ and $w_{ij}$ denote its estimated mean repair time. If no repair is needed on line $(i,j)$, both the reward and repair time parameters are set equal to zero. In a practical setting, time overruns are possible due to unforeseen eventualities and a repair which started in a previous restoration window and was scheduled for completion within that window may not actually have been finished. To accommodate such time overruns, we interpret $w_{ij}$ as the \emph{estimated mean residual repair time} of line $(i,j)$. In this paper, we concentrate on single source radial networks (more generally, acyclic graphs), but it is possible to extend our approach to any network with an arbitrary number of sources. We assume that a transportation graph, which is not shown in Figure~\ref{fig:Fig1_2}, is also available. This transportation graph allows us to determine the shortest travel times between all pairs of job sites. As mentioned in Section~\ref{sec:probDefn}, we assume that the travel times are symmetric and that it is possible to reach any site from any other site. In order to embed the repair times (associated with the distribution network) and the travel times (associated with the transportation network) within the same graph, we proceed as follows:
%
\begin{itemize}
  \item \emph{Step-$1$}: Contract the already energized portion of the network into a supernode (root), add an auxiliary node (node $0$), and add an undirected edge between the auxiliary node and the root. Assign to this edge the parameter vector $\left(w_{0r}=0, r_{0r}=0\right)$. The modified graph is denoted by $G_m$, as shown in Figure~\ref{fig:Fig1_2}(b).
  \item \emph{Step-$2$}: Construct the line graph of $G_m$, which we denote by $L(G_m)$. Note that the job sites appear as nodes in $L(G_m)$. Furthermore, each node in $L(G_m)$ is endowed with two parameters, a reward parameter and a repair time parameter, as shown in Figure~\ref{fig:Fig1_2}(c). Both these parameters are zero if a node does not need repair.
  \item \emph{Step-$3$}: Convert the line graph $L(G_m)$ to a complete graph, which we denote by $L_K(G_m)$; $L(G_m) \subseteq L_K(G_m)$. This step is shown in Figure~\ref{fig:Fig1_2}(d). We will refer to the node labeled $(0,r)$, interpreted as a dummy job, as the \emph{root} of $L_K(G_m)$. The repair time and reward parameters associated with each node in $L(G_m)$ carry over to $L_K(G_m)$. We will now assign weights to each undirected edge in $L_K(G_m)$. Let $c_{ij} = c_{ji}$, representing the shortest travel time between nodes $i$ and $j$, denote the weight of the edge $(i,j) \in L_K(G_m)$. For example, the weight of the undirected edge between the nodes $(4,8)$ and $(7,9)$ in Figure~\ref{fig:Fig1_2}(d) represents the travel time for a crew moving from the line/job $(4,8)$ in Figure~\ref{fig:Fig1_2}(a) to the line/job $(7,9)$ or \emph{vice versa}. Travel times between node $(0,r)$ and any other node are set to $0$, the reason for which will be apparent shortly.
\end{itemize}

A couple of observations are in order here. First, expansion of $L(G_m)$ to $L_K(G_m)$ in Step-$3$ allows the repair times, which appear as node weights in Figure~\ref{fig:Fig1_2}(d), and the travel times, which appear as edge weights in Figure~\ref{fig:Fig1_2}(d), to be embedded within the same \emph{undirected doubly weighted graph}. If the number of damaged edges in Figure~\ref{fig:Fig1_2}(a) is $|L^D|$, we end up with a doubly weighted complete graph in Figure~\ref{fig:Fig1_2}(d) on $|L^D|+1$ nodes. Second, while the presence of each edge in $L(G_m)$ reflects the physical connectivity of the underlying power distribution network $G$ (e.g., an edge exists between nodes $(r,4)$ and $(4,8)$ in Figure~\ref{fig:Fig1_2}(c) because the undirected edges $(r,4)$ and $(4,8)$ share a node in Figure~\ref{fig:Fig1_2}(b)), the interpretation of each edge in $L_K(G_m)$ is the time needed to travel the shortest path between the end nodes of that edge. This observation will be important when we discuss the nature of feasible solutions next.

The optimal repair schedules can be interpreted in the context of the traveling salesman problem. Assume that the number of repair crews is $m=1$ (one `repair salesman'). Operating within a certain time budget, the repair salesman starts from the node/home city $(0,r)$ in Figure~\ref{fig:Fig1_2}(d) and selectively chooses to visit certain cities (repair sites) such that the reward collected at the end of the tour is maximized. If the salesman decides to visit a certain city, he must spend a prescribed amount of time at that city, equal to the repair time of that job site, before moving to the next city following the shortest path between the two cities. Since the travel time from the home city to any other node is $0$, our repair salesman does not incur any cost while traveling to the first repair site, which is consistent with the assumption we made in Section~\ref{sec:probDefn} that any restoration window commences when all repair crew are at their first job sites. The total cost incurred by the salesman on his tour is the sum of the repair times (node weights in $L_K(G_m)$) and travel times (edge weights in $L_K(G_m)$) and the aggregate reward is the sum of the individual rewards collected at the cities visited. Since every non-root node in $L_K(G_m)$ is connected to the root by an edge with weight (travel time) $0$, the salesman can make his way back to his home city if visiting an additional city would result in a violation of the time budget, $\text{min}\left(\mathcal{T},\Delta T_k\right)$, where $\mathcal{T}$ represents the crew's time budget and $\Delta T_k := T_k - T_{k-1}$ denotes the length of the current restoration window.

\begin{figure*}[t]
  \centering
	\includegraphics[width=\textwidth]{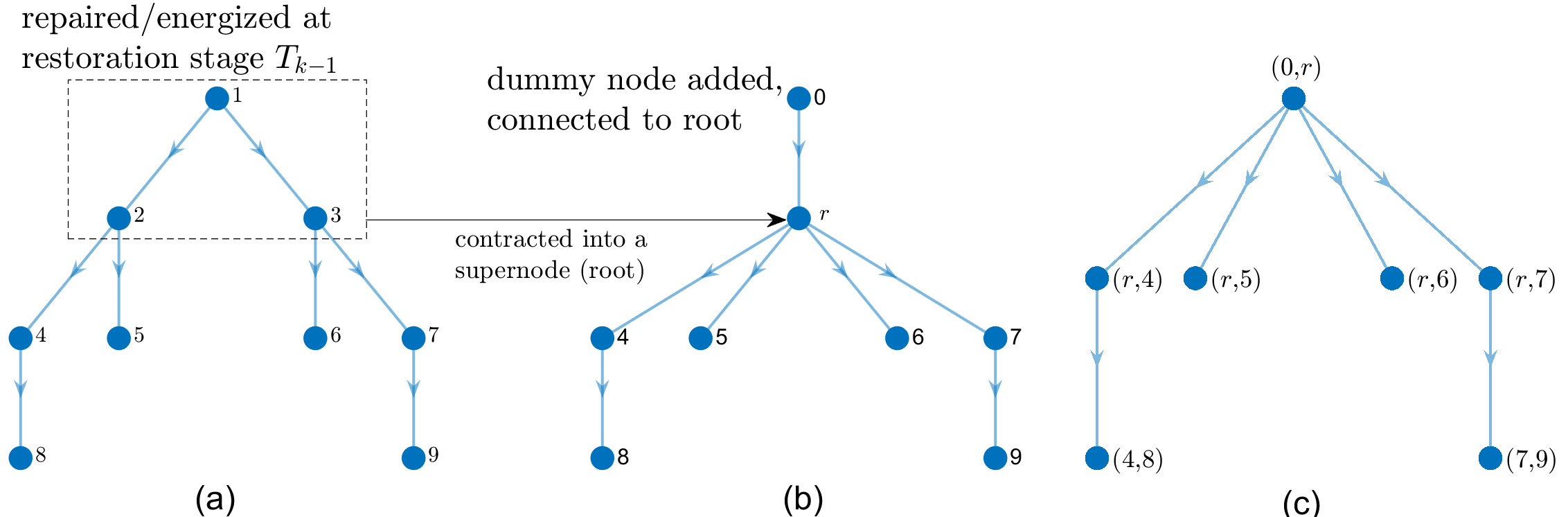}
	\caption{(a) Directed version of the radial distribution network shown in Figure~\ref{fig:Fig1_2}, which we denote by $G^d$. Arrow directions on the edges represent the direction of power flow. (b) Modified version of graph $G^d$, which we denote by $G_m^d$. (c) Line graph of the directed graph shown in panel (b), which we denote by $L(G_m^d)$.}
\label{fig:Fig5}
\end{figure*}

Three sets of constraints affect the solution of this variant of the TSP problem: (i) a time budget constraint for each crew which guarantees that the length (i.e., the aggregate of the repair times and the travel times) of each tour does not exceed that crew's time budget or the duration of the restoration window, whichever is smaller, (ii) a subtour elimination (or path continuity) constraint for each crew which guarantees that each crew solution is ``a tour'', and (iii)  electrical continuity constraints which guarantee that all nodes repaired during the $k^{\text{th}}$ restoration window can be connected to the already energized portion of the network (at the end of the $(k-1)^{\text{th}}$ restoration window. 

Electrical continuity constraints depend on the topology of the power distribution network. For single-source radial distribution networks, such constraints translate to \emph{precedence relations between the set of damaged nodes}. Consider the power distribution network shown in Figure~\ref{fig:Fig5}(a), which is the same network shown in Figure~\ref{fig:Fig1_2}(a), except that arrows have been added to indicate the direction of power flows. We denote this directed graph by $G^d$. First, we contract the repaired/energized portion in $G^d$ into a supernode, $r$, and add an auxiliary node, $0$, connected \emph{to} node $r$, as shown in Figure~\ref{fig:Fig5}(b). We denote this modified graph by $G_m^d$. Next, we convert $G_m^d$ to its line graph representation, $L(G_m^d)$, as shown in Figure~\ref{fig:Fig5}(c). This directed line graph captures all the precedence relations that the solution must satisfy. For example, if node $(7,9)$ is chosen to be repaired during the current window, we require that node $(r,7)$ be repaired first. Similarly, if node $(4,8)$ is chosen to be repaired during the current window, node $(r,4)$ must be repaired first. No precedence relations are required for nodes $(r,4)$, $(r,5)$, $(r,6)$, and $(r,7)$. It is easy to verify that if these precedence relations are enforced, the solution $(0,r) \to (r,4) \to (4,8) \to (7,9) \to (0,r)$, which violates the electrical continuity constraint, will be infeasible.
\subsection{Electrical continuity constraints for partially damaged networks}
\label{sec:partiallyDamaged}
In the previous paragraph, we assumed that all nodes in Figure~\ref{fig:Fig5}(c), other than the root node $(0,r)$, are damaged. For partially damaged scenarios, we adopt one more pre-processing step whereby the line graph $L(G_m^d)$ is ``collapsed'' to eliminate all non-damaged nodes. In order to avoid additional notation, we will continue to refer to the collapsed graph as $L(G_m^d)$, i.e., $L(G_m^d) \to \text{collapse procedure} \to L(G_m^d)$. Precedence relations are extracted as part of this procedure. 

\begin{figure}[htb]
  \centering
	\includegraphics[width=0.5\textwidth]{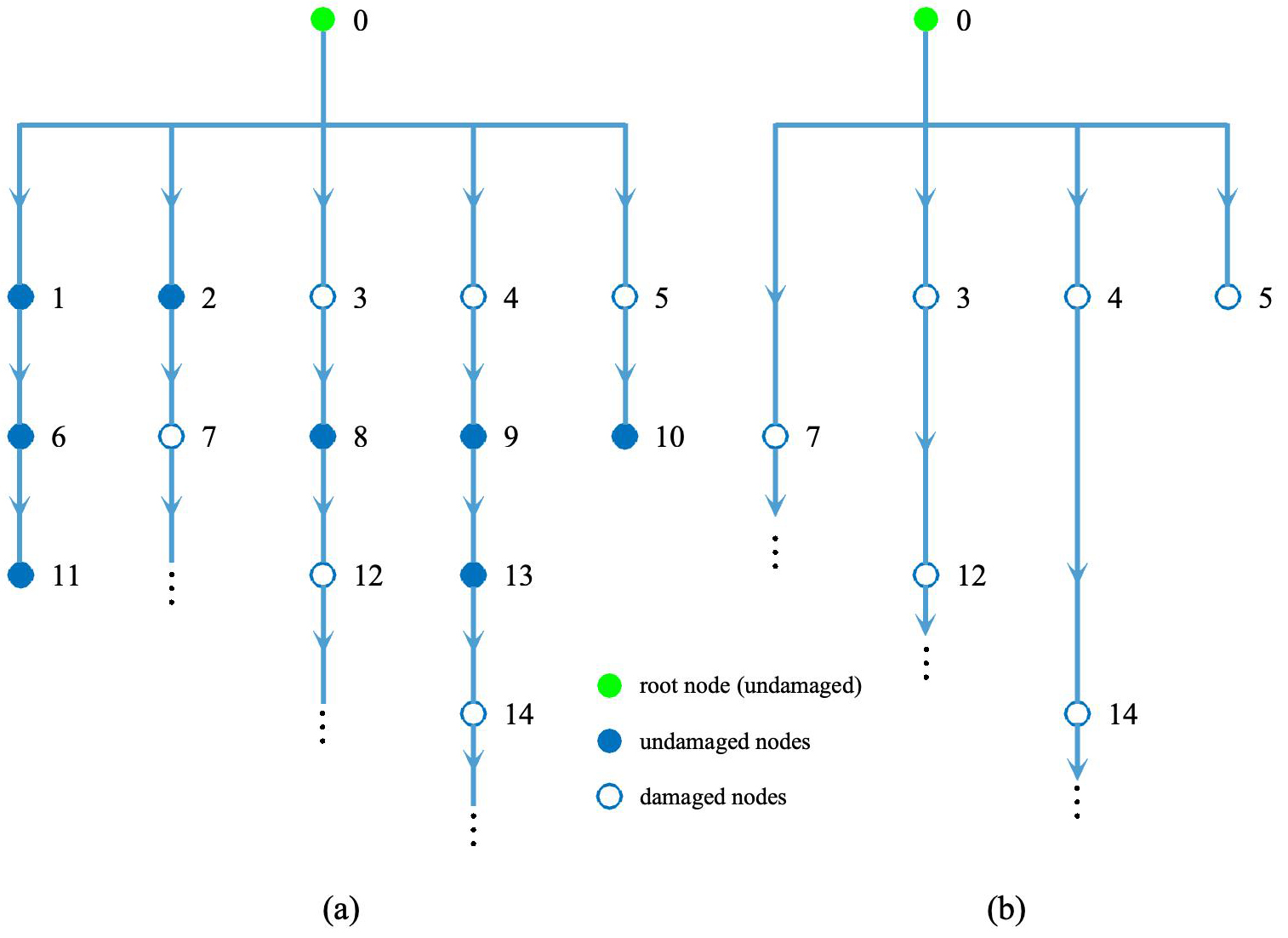}
	\caption{Illustrating the ``graph collapse'' procedure for a partially damaged network. (a) $L(G_m^d)$ as obtained after application of the procedure illustrated in Figure~\ref{fig:Fig5}. (b) $L(G_m^d)$ after the undamaged nodes (other than the root) are deleted.}
\label{fig:prefix}
\end{figure}

Consider the directed graph $L(G_m^d)$ shown in Figure~\ref{fig:prefix}(a), where we have adopted single-index node labels for simplicity, unlike the node labels in Figure~\ref{fig:Fig5}. Figure~\ref{fig:prefix}(b) shows the modified $L(G_m^d)$, which is obtained as follows: 
\begin{itemize}
    \item \emph{Step-$1$}: First, all undamaged leaf nodes in $L(G_m^d)$ are deleted, along with the directed edges incident on these nodes. Deletion of an undamaged leaf node may create an additional undamaged leaf node, which is also deleted. At the end of this step, there must not be any undamaged leaf node in $L(G_m^d)$. Executing this step on Figure~\ref{fig:prefix}(a) leads to the deletion of nodes labeled 10, 11, 6, and 1. Note that the last three nodes are deleted sequentially (after node 11 is deleted, node 6 becomes a leaf node, which is deleted next, followed by node 1).
    
    \item \emph{Step-$2$}: Next, for every damaged node $j \in L(G_m^d)$, we search for its nearest ancestor which is also damaged. If no such node is found, we record its nearest ancestor to be the root node (which, by definition, is undamaged). The precedence relation for node $j$ is therefore of the form $i \to j$, where $i$ is the nearest ancestor of $j$ which is also damaged. If $i = \emptyset$, we assign $i := \text{root node of $L(G_m^d)$}$. Executing this step on Figure~\ref{fig:prefix}(a) leads to the following precedence relations: (i) $0 \to 7$ (node 2 deleted), (ii) $0 \to 3$, (iii) $3 \to 12$ (node 8 deleted), (iv) $0 \to 4$, (v) $4 \to 14$ (nodes 9 and 13 deleted), and (ii) $0 \to 5$. These precedence relations are embodied in Figure~\ref{fig:prefix}(b). 
\end{itemize}
%
%

The previous discussion leads to an interesting question: `for a partially damaged network, how should one assign the rewards for a damaged node?' For example, referring to Figure~\ref{fig:prefix}(a), we see that if node 4 is repaired and energized, it \emph{may be} possible to energize nodes 9 and 13 simultaneously, provided there is a switch between nodes 13 and 14. In this case, the true reward for repairing node 4 is the sum of the individual rewards associated with nodes 4, 9, and 13. Accounting for node rewards in this manner, aided by knowledge of switch locations, may lead to improved decision making. A similar argument in favor of switch location guided repair scheduling can also be made for a fully damaged network. We plan on investigating this issue in future.

\begin{remark}
   In order to keep the set of nodes in $L_K(G_m)$ (see Figure~\ref{fig:Fig1_2}(d)) and $L(G_m^d)$ identical, nodes which are deleted as part of \emph{Step-$2$} above should also be deleted (along with incident edges) from $L_K(G_m)$.
\label{remark:partiallyDamaged}
\end{remark}


\section{MILP formulation}
\label{sec:radialGraphs}
While we have cycled through different graph representations, we will define our MILP formulation of the $m$-crew repair scheduling problem on the undirected complete graph $L_K(G_m)$ (see Figure~\ref{fig:Fig1_2}(d)) and the directed graph $L(G_m^d)$ (see Figure~\ref{fig:Fig5}(c)). We will use $L_K(G_m)$ to set up the repair scheduling problem as a \emph{traveling salesman-like} problem and $L(G_m^d)$ to impose the additional electrical continuity constraints. For the rest of the paper, we will refer to these two graphs as \emph{working graphs} and denote them by $G_w$ and $G_{wd}$ respectively, the subscript $w$ indicating `working' and the subscript $d$ indicating `directed'. Variables defined on undirected edges will bear the subscript $(ij)$; e.g., $x_{(ij)c}=1$ means that the undirected edge \emph{between} $i$ \emph{and} $j$ is traveled by crew $c$. Variables defined on directed edges bear the subscript $ij$; e.g., $x_{ijc}=1$ means that the directed edge \emph{from} $i$ \emph{to} $j$ is traveled by crew $c$. Table~\ref{tab:notationTable} summarizes the major notation used in the rest of the paper. 
%
%
\begin{table}[t]
    \renewcommand{\arraystretch}{1.5}
    \caption{ Notations used in the MILP model}
    \label{tab:notationTable}
    \centering

    \begin{tabular}{| p{0.12\linewidth} | p{0.78\linewidth} |} \hline
      \hspace*{0ex}\textbf{Notation} & \hspace*{5ex} \textbf{Description} \\ \hline \hline
      $G_w$ & undirected complete working graph, $L_K(G_m)$ (see Figure~\ref{fig:Fig1_2}(d)). \\ \hline
      $\mathcal{E}_w$ & set of undirected edges in $G_w$. \\ \hline
      $G_{wd}$ & directed working graph, $L(G_m^d)$ (see Figure~\ref{fig:Fig5}(c)).  \\ \hline
      $\mathcal{E}_{wd}$ & set of directed edges in $G_{wd}$. \\ \hline
      $\mathcal{N}_w$ & set of all nodes in $G_w$ or $G_{wd}$; $\mathcal{N}_w = \{0,1,2,\ldots n-1\}$, where node $0$ is the root/home city.  \\ \hline
      $n$ & number of nodes in $G_w$ or $G_{wd}$ \\ \hline
      $\mathcal{N}_w^D$ & subset of nodes in $\mathcal{N}_w$ which are damaged and require repair; note that node 0 is not in $\mathcal{N}_w^D$. \\ \hline 
      $r_i$ $(\geq 0)$ & reward of node $i \in \mathcal{N}_w$; $r_i = 0$, $\forall i \in \left\{\mathcal{N}_w \setminus \mathcal{N}_w^D\right\}$. \\ \hline
      $w_i$ & repair time of node $i \in \mathcal{N}_w$; $w_i = 0$, $\forall i \in \left\{\mathcal{N}_w \setminus \mathcal{N}_w^D\right\}$. \\ \hline
      $c_{(ij)}$ & travel time \emph{between} nodes $i,j \in \mathcal{N}_w$, following the shortest path between $i$ and $j$; i.e., $c_{(ij)} = c_{ij} = c_{ji}$. If either $i=0$ or $j=0$, then $c_{(ij)} = 0$  \\ \hline
      $\mathcal{C}$ & set of crew indices; $\mathcal{C} = \{1,2,\ldots m\}$ \\ \hline
      $m$ & number of repair crew (repair salesmen) \\ \hline
      $\mathcal{T}_c$ & time budget of crew $c$ \\ \hline
      $\Delta T_k$ & time duration of the current ($k^{th}$) restoration window. \\ \hline
      $y_{ic}$ & binary variable which takes the value $1$ if crew $c$ visits/repairs node $i \in \mathcal{N}_w^D$ or node $i$ does not need repair, i.e., $i \in \left\{\mathcal{N}_w \setminus \left\{ \mathcal{N}_w^D \cup 0 \right\}\right\}$, $0$ otherwise. \\ \hline
      $x_{(ij)c}$ & integer variable which takes the value $0$ if crew $c$ does not travel between nodes $i,j \in \mathcal{N}_w$, $i \neq j$. If crew $c$ travels between nodes $i$ and $j$, then $x_{(ij)c} \in \{1,2\}$, provided the undirected edge $(i,j)$ is between node $0$ and some other node. If $(i,j)$ is not incident on the root, $x_{(ij)c} = 1$. \\ \hline
      $(i,j)$ & undirected edge between nodes $i$ and $j$. \\ \hline
      $i \to j$ & directed edge from node $i$ to $j$. \\ \hline
      $\{i,j\}$ & set of nodes $i$ and $j$. \\ \hline
    \end{tabular}
\end{table}

In this section, we discuss an MILP model for the repair scheduling problem on a radial power distribution network with a single source. Our assumption of a radial distribution network isn't too restrictive since, even if such networks are topologically non-radial (e.g., meshed), they are operated radially. If operational decisions are made prior to repair planning/scheduling, allowing the non-radial network to be pruned and pre-configured to a radial structure, the MILP model discussed in this paper will be applicable. 

Given a distribution network $G$, we first perform the pre-processing steps summarized below to create the two working graphs, $G_w$ and $G_{wd}$ (see Table~\ref{tab:notationTable}):
\begin{alignat*}{4}
  &G \, \to \, && \, G_m \, \to \, && L(G_m) \, \to \, \overbrace{L_K(G_m)}^{\text{Figure~\ref{fig:Fig1_2}(d)}} &&:= G_w = \left(\mathcal{N}_w,\mathcal{E}_w\right) \\
  &\downarrow  && && && \\
  &G^d \, \to \, && \, G_m^d \to && \underbrace{L(G_m^d)}_{\text{Figure~\ref{fig:Fig5}(c)}}  &&:= G_{wd} = \left(\mathcal{N}_w,\mathcal{E}_{wd}\right)
\end{alignat*}
Without any loss of generality, we assume that all nodes in $G_{w}$ and $G_{wd}$, other than the root node, are damaged. For a partially damaged scenario, this requires the ``graph collapse'' procedure explained in Section~\ref{sec:partiallyDamaged} to be implemented (see also Remark~\ref{remark:partiallyDamaged}). As mentioned previously, we use $G_w$ to set up the repair scheduling problem as a \emph{traveling salesman-like} problem and $G_{wd}$ to impose the additional electrical continuity constraints. Note that $\left|\mathcal{E}_w\right| = n(n-1)/2$ since $G_w$ is an undirected graph. Accordingly, we choose the set of edges in $G_w$ to be $\mathcal{E}_w = \{(i,j): 0 \leq i \leq n-2, \, j > i\}$. Our decision variables are $\left\{y_{ic}: i \in \mathcal{N}_w, \, c \in \mathcal{C}\right\}$ and $\left\{x_{(ij)c}: (i,j) \in \mathcal{E}_w, \, c \in \mathcal{C}\right\}$. These variables are defined in Table~\ref{tab:notationTable}.

The MILP formulation for $m$-crew repair scheduling is listed in Figure~\ref{fig:ILP_model_1}. We dispatch $m$ `repair salesmen', each salesman starting and ending his tour at the root node $0$. Additionally, each salesman operates under his own time budget and each node which needs repair can be visited by at most one repair salesman. Our objective  (eqn.~(\ref{eq:objFn})) is to maximize the aggregate reward that can be collected by all crews at the end of the current restoration window. If $r_i = 1$, for all $i \in \mathcal{N}_w^D$, the objective reduces to maximizing the number of nodes that can be repaired within the current restoration window. Adding the penalty term $- \sum_{i \, \in \, \mathcal{N}_w^D} (1 - \sum_{c \, \in \, \mathcal{C}} y_{ic}) \, p_i$ to the cost function, where $p_i$ is the penalty parameter for the $i^{\text{th}}$ repair job in the current restoration window (in principle, $p_i$ can be a function of time, without or with saturation as in a sigmoid, to reflect the aggregate time that the $i^{\text{th}}$ job has been backlogged for) changes the objective from reward maximization to profit maximization and introduces a notion of \emph{fairness} to the recovery schedule. 

The roles of the constraint equations are discussed below.
%
%
\begin{itemize} 
  \item Eqn.~(\ref{eq:const1}) forces each crew to visit the root (node $0$).
  \item Eqn.~(\ref{eq:const2}) imposes binary conditions on all nodes in $\mathcal{N}_w^D$ (set of all nodes which need repair), on a per crew basis.
  %
  %
  \item Eqns.~(\ref{eq:const5}) and (\ref{eq:const6}) impose integrality conditions on the triple indexed $x_{(ij)c}$ variables. The set of values these variables can take depends on whether the undirected edge $(i,j)$ is incident on the root node or not. If $(i,j)$ is not incident on the root, $x_{(ij)c}$ is binary valued (eqn.~(\ref{eq:const6})). In contrast, if $(i,j)$ is incident on the root, $x_{(ij)c}$ takes a value from the set $\{0,1,2\}$ (eqn.~(\ref{eq:const5})). We allow $x_{(ij)c}$ to take the value $2$ in the latter case since we accept cycles of length $2$ in the optimal solution. For example, referring to Figure~\ref{fig:Fig1_2}(d), the optimal solution for $m=2$ could be $(0,r) \to (r,4) \to (0,r)$ and $(0,r) \to (r,5) \to (r,6) \to (0,r)$. In this case, the undirected edge between $(0,r)$ and $(r,4)$ is used twice by the first crew, but all edges traveled by the second crew are used once. 
  \item Eqn.~(\ref{eq:const7}) stipulates that all nodes needing repair be visited/repaired by at most one crew.
  \item Eqn.~(\ref{eq:const8}) relates the $x$ variables to the $y$ variables, on a per crew basis. If crew $c$ visits node $i$, where $i$ is either the root node or belongs to the set of nodes needing repair, then $y_{ic} = 1$, which implies $\sum_{j : (ij) \, \in \, \mathcal{E}_w}  x_{(ij)c} = 2$. This leaves open two possibilities: (i) some undirected edge incident on node $i$ is used twice by the crew, which happens only when the edge connects to the root and the optimal solution for that crew is a cycle of length $2$ (i.e., the tour is of the form $0 \to i \to 0$), or (ii) two different edges incident on $i$ are used, once each, which happens when the optimal solution for that crew is a cycle of length greater than $2$, in which case one edge is used to enter the node and another to exit (i.e., the tour is of the form $0 \to i \to q \cdots \to 0$ or $0 \to \cdots \to p \to i \to q \cdots \to 0$). 
  
  Eqn.~(\ref{eq:const8}) can sometimes produce counter-intuitive results with respect to the number of crews $m$, in the sense that the problem can be feasible for some values of $m$ and $m+2$, but infeasible for $m+1$. Sec.~V(A) of \cite{wei2024rolling} provides a detailed discussion and discusses possible solution strategies. For this paper, whenever we encountered such infeasibilities during simulations, we discarded those problem instances. 
  %
  %
  \item Eqn.~(\ref{eq:const10}) imposes a time constraint on every tour, on a per crew basis. For crew index $c$, the maximum tour cost (in units of time) is the minimum of $\mathcal{T}_c$ and $\Delta T_k$, where $\mathcal{T}_c$ is the time budget for crew $c$ and $\Delta T_k$ is the time duration of the $k^{th}$ restoration window. Since we do not allow fractional repairs (and therefore, prorated rewards), the l.h.s of eqn.~(\ref{eq:const10}) may not equal the r.h.s in the optimal solution. Note that both repair times and travel times are accounted for in the l.h.s of eqn.~(\ref{eq:const10}). 
  \item Eqn.~(\ref{eq:const11}) enforces the electrical continuity constraint by ensuring that for every node which is visited/repaired by \emph{some} crew, \emph{a directed fully repaired path exists from the source to that node}. This is achieved by imposing the complete set of precedence relations dictated by the working graph $G_{wd}$ (i.e., $L(G_m^d)$; see Figure~\ref{fig:Fig5}(c),  Figure~\ref{fig:prefix} for a partially damaged network, and Remark~\ref{remark:partiallyDamaged}). 
  %
  
  In essence, the electrical continuity constraints yield a pattern of energized swathes which expand radially outward from the source over time, which makes intuitive sense for single-source radial distribution networks. For other topologies with either single or multiple sources, these constraints may need to be modeled differently.
  \item Eqn.~(\ref{eq:const12}) is a placeholder for subtour elimination constraints (SECs) which we address below.
\end{itemize}

\begin{figure}[t]
%
\begin{minipage}{\linewidth}
  \centering
  \rule{\linewidth}{1pt}
\end{minipage}\hfill%
\begin{subequations}
\begin{alignat}{3}
  \text{max } & \sum_{i \, \in \, \mathcal{N}_w^D} r_i \, \sum_{c \, \in \, \mathcal{C}} y_{ic} && && \label{eq:objFn} \\
  \text{s.t} \quad & && && \nonumber \\
  & y_{ic} = 1; \ \forall c \in \mathcal{C}, \ i = 0 \label{eq:const1} \\
  & y_{ic} \in \{0,1\}; \ \forall c \in \mathcal{C}, \ \forall i \in \mathcal{N}_w^D \label{eq:const2} \\
  %
  %
  %
  & x_{(0j)c} \in \{0,1,2\}; \ \forall c \in \mathcal{C}, \ \forall (i,j) \, \in \, \mathcal{E}_w,  \label{eq:const5} \\
  & x_{(ij)c} \in \{0,1\}; \ \forall c \in \mathcal{C}, \ \forall (i,j) \, \in \, \mathcal{E}_w, \ i \neq 0 \label{eq:const6} \\
  \sum_{c \, \in \, \mathcal{C}} & y_{ic} \leq 1; \ \forall i \in \mathcal{N}_w^D \quad (\text{note: } 0 \not\in \mathcal{N}_w^D)\label{eq:const7} \\
  %
  \sum_{j : (ij) \, \in \, \mathcal{E}_w} & x_{(ij)c} - 2 \, y_{ic} = 0; \ \forall c \in \mathcal{C}, \ \forall i \in \left\{\mathcal{N}_w^D \cup 0\right\} 
  \label{eq:const8} \\
  %
  \label{eq:const9} \\
  %
  \sum_{i \, \in \, \mathcal{N}_w^D} & w_i \, y_{ic} + \sum_{(i,j) \, \in \, \mathcal{E}_w} c_{(ij)} \, x_{(ij)c} \leq \text{min}\left(\mathcal{T}_c, \Delta T_k\right); \forall c \in \mathcal{C} \label{eq:const10} \\
  %
 \sum_{c \, \in \, \mathcal{C}} & y_{ic} - \sum_{c \, \in \, \mathcal{C}} y_{jc} \geq 0; \ \forall (i \to j) \in \mathcal{E}_{wd}
  \label{eq:const11} \\
  %
  & \text{subtour elimination constraints} && && \label{eq:const12}
\end{alignat}
\label{eq:ILPmdl}
\end{subequations}
\captionsetup[subfigure]{labelformat=empty}
\begin{minipage}{\linewidth}
  \centering
  \rule{\linewidth}{1pt}
\end{minipage}\hfill%
\caption{MILP formulation for $m$-crew repair scheduling during  the $k^{th}$ restoration window. Note that the electrical continuity constraints, eqn.~(\ref{eq:const11}), are valid only for single source, radial power distribution networks.}
\label{fig:ILP_model_1}
\end{figure}

\subsection{Subtour elimination constraints}
\label{sec:radialGraphs_SEC}
There exists a rich body of literature on different approaches for subtour elimination in the traveling salesman problem. Some of these approaches require an exponential number of constraints, while others are polynomial formulations. 
In this paper we adopt and modify the single commodity flow (SCF) based subtour elimination procedure proposed by Gavish and Graves \cite{rGavishGraves1978}. In the context of the TSP, the SCF-based subtour elimination procedure requires that the home city be able to send a unit of flow to each of the cities visited, under the constraint that a positive flow can be activated on the edge $i \to j$ only if the salesman has traveled from $i$ to $j$. Clearly, for the flow constraints to be satisfied, there must not be any subtour in the optimal solution.

To accommodate a flow model, we view each undirected edge $(i,j) \in \mathcal{E}_w$ as a bidirected edge with corresponding continuous-valued and non-negative flow variables $f_{ij}$ and $f_{ji}$, where $f_{ij}$ is interpreted as a flow \emph{from} $i$ \emph{to} $j$. We define the `flow differential' at any node as the aggregate flow out of the node minus the aggregate flow into the node. Our SCF based subtour elimination formulation is shown in eqn.~(\ref{eq:SCFbasedSEC}), which requires additional $O(n^2)$ continuous variables and $O(n^2)$ constraints. Notice that all inequalities are coupling constraints, in the sense that every equality/inequality depends on the collective job completions by all $m$ crews. 

Eqn.~(\ref{eq:SCFbasedSEC_1}) ensures that the flow differential at the root is equal to the total number of jobs completed by all crew. Eqn.~(\ref{eq:SCFbasedSEC_2}) ensures that the flow differential at any node which needs repair is equal to $-1$ if it has been visited/repaired by any crew, and $0$ otherwise. Eqn.~(\ref{eq:SCFbasedSEC_3}) is a flow activation condition which ensures that flows on $i \to j$ and $j \to i$ can be activated only if some crew has traveled the undirected edge $(i,j)$, i.e., $\sum_{c \, \in \, \mathcal{C}} x_{(ij)c} \geq 1$. If flows are activated, it also imposes a capacity constraint which limits the maximum flow to either $2mn$ or $2n^2$. To see why, consider some node $i \in \left\{\mathcal{N}_w^D \cup 0\right\}$. Summing eqn.~(\ref{eq:const8}) over $c$, we find:
\begin{align}
   \sum_{c \, \in \, \mathcal{C}} \sum_{j : (ij) \, \in \, \mathcal{E}_w} x_{(ij)c} \leq 2 \sum_{c \, \in \, \mathcal{C}} y_{ic}, \ \forall i \in \left\{\mathcal{N}_w^D \cup 0\right\}
\label{eq:const8summed}
\end{align}
If $i = 0$, the r.h.s of eqn.~(\ref{eq:const8summed}) evaluates to $2m$, using eqn.~(\ref{eq:const1}). Since:
\[ \sum_{c \, \in \, \mathcal{C}} \sum_{j : (ij) \, \in \, \mathcal{E}_w} x_{(ij)c} \leq 2m \ \Rightarrow \ \sum_{c \, \in \, \mathcal{C}} x_{(ij)c} \leq 2m, \]
eqn.~(\ref{eq:SCFbasedSEC_3}) reduces to:
\[ f_{ij}, \, f_{ji} \leq n \sum_{c \, \in \, \mathcal{C}} x_{(ij)c} = 2mn \]
However, if $i \in \mathcal{N}_w^D$, the r.h.s of eqn.~(\ref{eq:const8summed}) evaluates to $2n$, using eqn.~(\ref{eq:const7}), and eqn.~(\ref{eq:SCFbasedSEC_3}) reduces to:
\[ f_{ij}, \, f_{ji} \leq n \sum_{c \, \in \, \mathcal{C}} x_{(ij)c} = 2n^2 \]
%
%
Eqn.~(\ref{eq:SCFbasedSEC_6}) significantly tightens the flow upper bound established by eqn.~(\ref{eq:SCFbasedSEC_3}) by restricting the maximum flow on any directed edge to the total number of nodes visited/repaired by all crew, which can be at most equal to the cardinality of $\mathcal{N}_w^D$ (i.e., number of nodes needing repair). 
%
%
\begin{figure}[t]
\captionsetup[subfigure]{labelformat=empty}
\begin{minipage}{\linewidth}
  \centering
  \rule{\linewidth}{1pt}
\end{minipage}\hfill%
\begin{subequations}
\begin{alignat}{2}
   \sum_{j=1}^n f_{0j} - \sum_{j=1}^n f_{j0} &= \sum_{c \, \in \, \mathcal{C}} \, \sum_{i \, \in \, \mathcal{N}_w^D} y_{ic}  && \label{eq:SCFbasedSEC_1} \\
   \sum_{\substack{j=0 \\ j\neq i}}^n f_{ij} - \sum_{\substack{j=0 \\ j\neq i}}^n f_{ji} &= - \sum_{c \, \in \, \mathcal{C}} y_{ic}; \hspace{1.35cm} \forall i \in \mathcal{N}_w^D \label{eq:SCFbasedSEC_2} \\
   f_{ij}, \, f_{ji} &\leq n \sum_{c \, \in \, \mathcal{C}} x_{(ij)c}; \hspace{1cm} \forall (i,j) \in \mathcal{E}_w \label{eq:SCFbasedSEC_3}  \\
   f_{ij}, \, f_{ji} &\leq \sum_{c \, \in \, \mathcal{C}} \sum_{i \, \in \, \mathcal{N}_w^D} y_{ic};  \hspace{0.75cm} \forall (i, j) \in \mathcal{E}_w \label{eq:SCFbasedSEC_6} \\
   f_{ij}, \, f_{ji} &\geq 0; \,  \hspace{2.55cm}\forall (i,j) \in \mathcal{E}_{w} \label{eq:SCFbasedSEC_7}
\end{alignat}
\label{eq:SCFbasedSEC}
\end{subequations}
\captionsetup[subfigure]{labelformat=empty}
\begin{minipage}{\linewidth}
  \centering
  \rule{\linewidth}{1pt}
\end{minipage}\hfill%
\caption{Single commodity flow based subtour elimination constraints to be used in conjunction with the MILP model in Figure~\ref{fig:ILP_model_1}. Each undirected edge $(i,j)$ in $\mathcal{E}_w$ gives rise to two directed flow variables, $f_{ij}$ and $f_{ji}$.}
\label{fig:ILP_model_2}
\end{figure}
\section{Computational Results}
\label{sec:compResults}
In this section, we evaluate the base model shown in Figure~\ref{fig:ILP_model_1} on the IEEE 123-node test feeder network. The following modifications were made to the standard IEEE 123-node network (see Fig. A3 in \cite{wei2024rolling}):
\begin{itemize}
  \item Link distances in the standard IEEE test network \cite{IEEnetworks}, which are specified as 0 ft are set equal to 1 ft. 
  \item Since our model is designed exclusively for radial networks, links $(151, 300)$ and $(54, 94)$ were deleted from the IEEE network to make it radial. Additionally, the switches between nodes 151 and 300 and between nodes 54 and 94 were deemed to be in the open state. 
\end{itemize}
Our simulations were conducted using the Gurobi \cite{Gurobi} solver with the following parameters: 
\begin{itemize}
    \item all nodes other than the source are damaged and require repair, 
    \item the rewards for all jobs are equal to $1$, i.e., $r_i = 1, \, \forall i$, 
    \item job repair times were drawn from a Weibull distribution with varying means, \emph{subject to a lower bound of $30$ minutes} (consequently, all repair times are 30 minutes or longer), 
    \item we assume that the road infrastructure network is identical to the power network. Travel distance between job sites $i$ and $j$ is thus equal to the shortest path distance between $i$ and $j$, which is derived from the topology and length of the various segments of this network,
    \item the travel time between jobs $i$ and $j$ is equal to the shortest path distance between $i$ and $j$, divided by the travel speed,
    \item the travel speed was adjusted to yield varying travel time matrices,
    \item the time budgets for all crew are identical,
    \item node 150 was selected as the `home node' (node 0). 
\end{itemize}
Simulations were conducted on the University of Washington's Hyak computer cluster \cite{UWHyak}. For each simulation, we allocated 20 tasks per node and a total of 80 GB of RAM. We conducted our simulations with a prescribed termination criterion of 13\% duality gap (``MIPgap'' parameter in Gurobi) or 12 hours of solver time, whichever occurs first. Upon program termination, the mean and worst-case \emph{gap} (defined as $\text{UB - LB}$, where UB and LB denote the best feasible upper and lower bounds respectively) we observed over all trials turned out to be 2.05 (4.8\%) and 7 (12.96\%) respectively. We evaluate our model based on three performance metrics, described subsequently.  

\begin{figure}[t]
    \centering
    \includegraphics[width=\linewidth]{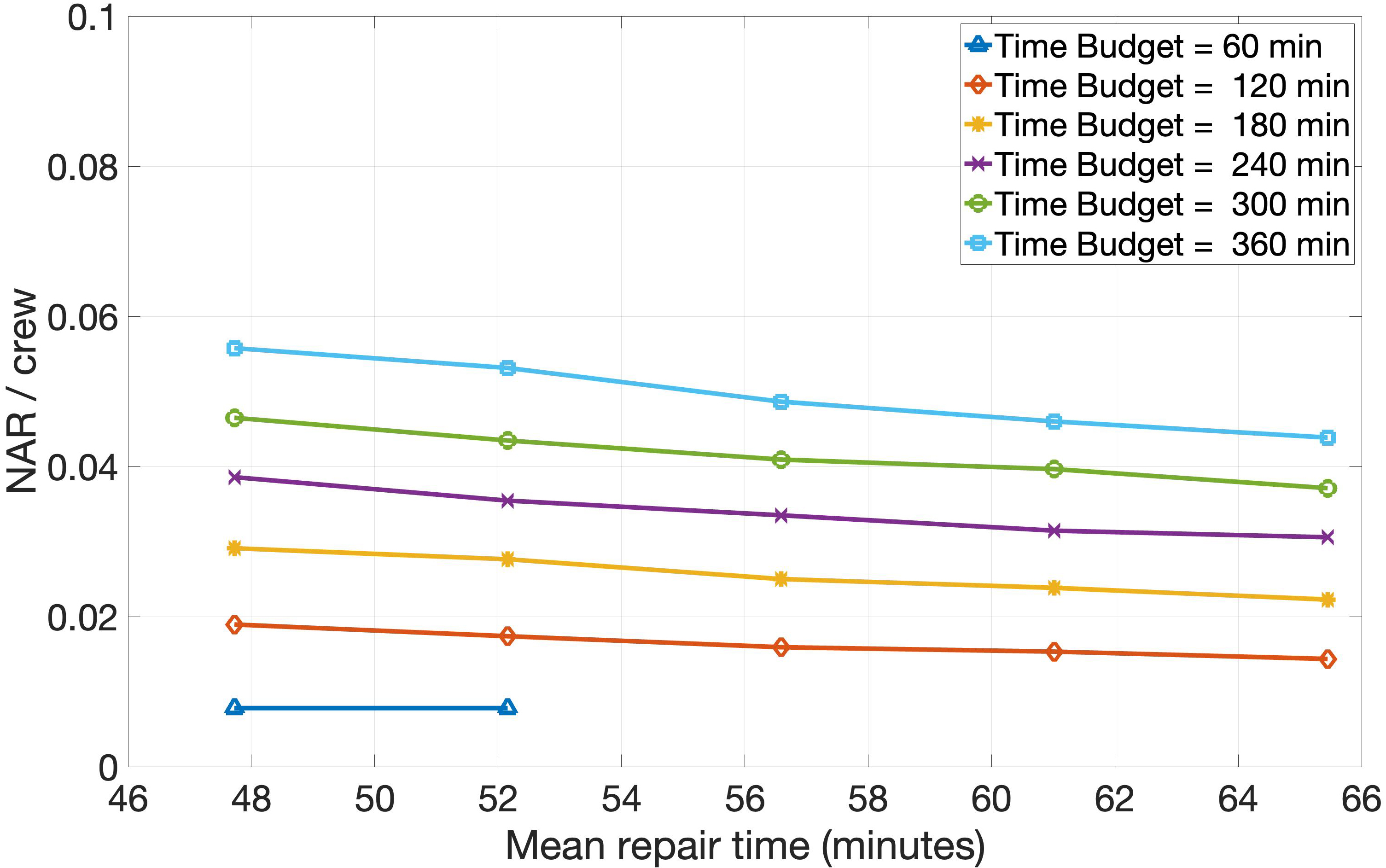}
    \caption{Normalized aggregate reward (NAR) per crew as a function of the mean repair time and time budget per crew for $m=8$.}
    \label{fig:AggRwd123}

    \includegraphics[width=\linewidth]{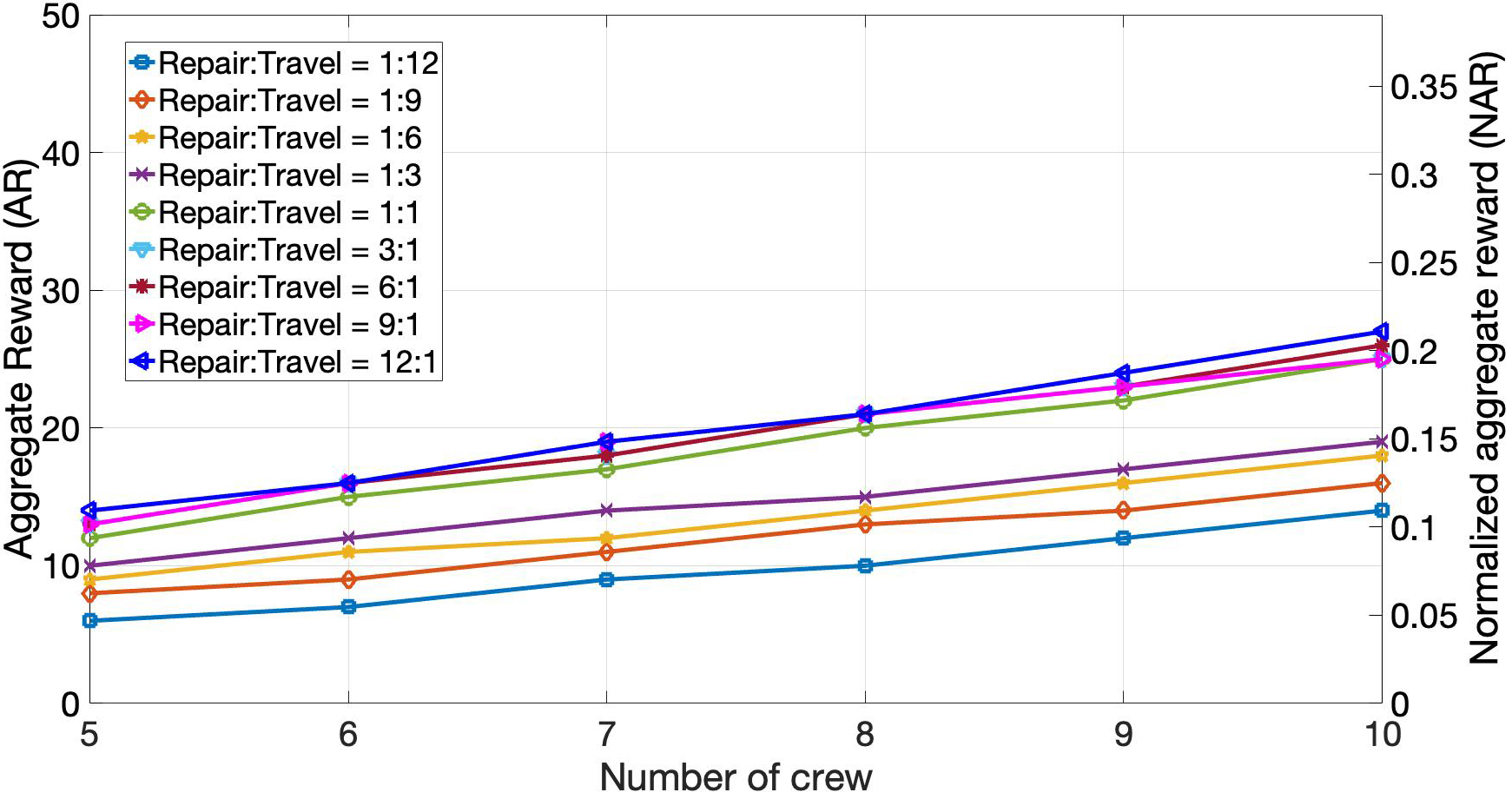}
    \caption{Aggregate reward (AR) and normalized aggregate reward (NAR) as a function of the number of crews and the mean repair to travel time ratio. The time budget for each crew is 180 minutes.}
    \label{fig:AggRwdCrews123}

    \includegraphics[width=0.95\linewidth]{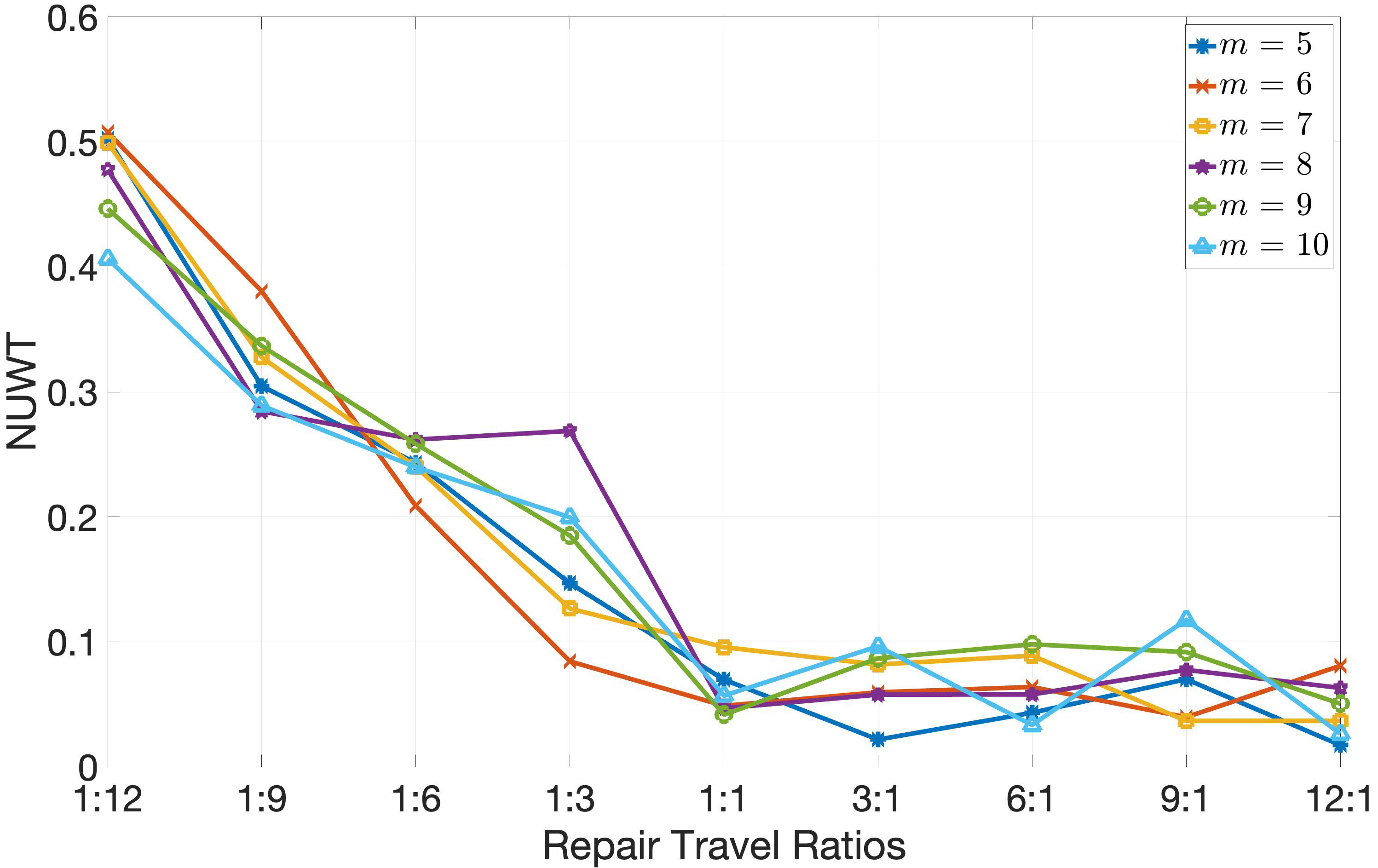}
    \caption{Normalized unused work time (NUWT) as a function of the mean repair to travel time ratio and the number of crew $m$. The time budget for each crew is 180 minutes.}
    \label{fig:NUWT123}
\end{figure}

We define the normalized aggregate reward (NAR) per crew as the aggregate reward (i.e, the value of the objective function upon program termination) divided by the product of the number of nodes ($n-1$) and the number of crews ($m$). Figure~\ref{fig:AggRwd123} shows the NAR per crew as a function of the mean repair time and time budget per crew, for $m=8$. Each data point in Figure~\ref{fig:AggRwd123} is obtained by averaging over $10$ trials. Observe that the maximum NAR per crew that can be achieved when $m=8$ is $1/8=0.125$. The travel speed was set such that the mean repair to travel time ratio is 3:1 when the mean repair time is 47.725 minutes. This required using a speed of 225 ft/min ($\approx 2.6$ mph). Although this speed is unrealistic, using this value in conjunction with IEEE-specified edge lengths resulted in realistic travel times, in the range of  $[0.025, 216.325]$ minutes. Keeping the travel speed (or equivalently, travel times) fixed at the above value, we then varied the mean repair time and the time budget/crew to generate the plots shown in Figure~\ref{fig:AggRwd123}. When the time budget/crew is very high (e.g., 240, 300, or 360 minutes) relative to the repair and travel times, the NAR/crew achieves its highest possible value (approximately 0.058) when the mean repair time is swept between $47-66$ minutes (approximately). Conversely, for very low time budgets (60 minutes), the NAR/crew remains constant at approximately $1/122 \approx 0.008$, which corresponds to 8 completed jobs out of a possible 122. For moderate time budgets, we expect the NAR/crew to be monotonically decreasing as the mean repair time increases while keeping the travel times fixed. This is indeed what we observe when the time budget per crew is 120 or 180 minutes. Additionally, when the time budget is very small (60 minutes) and the mean repair time exceeds approximately 56 minutes, no job meeting the electrical continuity constraints could be completed by any crew, which resulted in a NAR/crew of zero. This explains why the $x$-axis is limited to approximately 52 minutes when the time budget per crew is 60 minutes.

Figure~\ref{fig:AggRwdCrews123} shows the aggregate reward as a function of the number of crews and of the repair to travel time ratio. For convenience, the $y$-axes are dual scaled to show the aggregate reward as well as the normalized aggregate reward (NAR). Note that per crew normalization is not performed (unlike in Figure~\ref{fig:AggRwd123}) because our intent is to understand the marginal benefit stemming from the deployment of an additional crew. We fixed the repair time vectors (the mean is approximately 79 minutes) and varied the travel speed to obtain different travel time matrices/distributions. This resulted in different ratios of mean repair time to mean travel time. Since the travel speed and the repair time vectors are assumed deterministic, each data point in Figure~\ref{fig:AggRwdCrews123} is derived from a single trial.

%
%
%
 Intuitively, for a fixed repair to travel time ratio, we expect the aggregate reward to be non-decreasing, which is confirmed in Figure~\ref{fig:AggRwdCrews123}. Additionally, for a fixed repair time vector, a larger repair to travel time ratio (e.g., 12:1) implies smaller commute times which allows for more jobs to be completed (higher aggregate reward), while a smaller repair to travel time ratio (e.g., 1:12) implies larger commute times which results in fewer job completions (smaller aggregate reward). A visual examination of the color coded sequence of plots in Figure~\ref{fig:AggRwdCrews123} confirms this behavior.

Since fractional job completions are not allowed in our model (rewards earned on job completion, provided electrical continuity constraints are met, are therefore all-or-nothing), the total work time (i.e., the total time spent on repair and travel) for any crew need not be exactly equal to the time budget allocated to that crew. We refer to this residual time as the unused work time for that crew. There is another reason why the unused work time need not be zero. Depending on the number of crews deployed and the time budget for each crew, it may be possible for one or more crews to complete their repair schedule strictly before their respective time budgets. In this case too, the unused work time will be non-zero. We define the normalized unused work time (NUWT), viewed as measure of crew utilization, as the ratio of total unused work time over all crews divided by the total work time of all crews. Figure~\ref{fig:NUWT123} shows the NUWT as a function of mean repair to travel time ratio and number of crew. The repair and travel time parameters for Figure~\ref{fig:NUWT123} are identical to those used for Figure~\ref{fig:AggRwdCrews123}. As in Figure~\ref{fig:AggRwdCrews123}, each data point in Figure~\ref{fig:NUWT123} is derived from a single trial. For a fixed repair time vector, when the ratio of the number of available repair jobs to number of crew is high (recall that we assume that all but the source node is damaged) and the time budgets are reasonable, we can expect that most of the crews will be 'maximally utilized', i.e., be engaged in either repair or travel for a substantial fraction of their time budgets, particularly when travel times are small (or alternately, large mean repair to travel time ratios). This should result in small unused work time. For large travel times, it may not be possible for crews to make multiple commutes, possibly resulting in a larger unused work time. Figure~\ref{fig:NUWT123} confirms this behavior. We also observe that the unity repair to travel ratio is an elbow point of the family of NUWT curves.

\section{Conclusion} 
\label{sec:conclusion}
In this paper, we address the problem of optimally restoring electrical distribution networks after a severe weather event. We develop a new variant of the Traveling Salesman Problem (TSP) that incorporates various significant practical constraints into one comprehensive optimization problem. Unlike the majority of other approaches, which often neglect or simplify these practical constraints, this approach ensures that the restoration schedule is not only optimal but also practical. The core of our approach is the reconstitution of the damage graph into a doubly-weighted undirected graph. The optimal restoration scheduling problem can then be formulated and solved using Mixed Integer Linear Programming (MILP). Simulations on the IEEE 123-node test feeder show that our approach provides near-optimal solutions that consider these constraints on large networks that have suffered significant damage.

\bibliographystyle{IEEEtran}
\bibliography{IEEEabrv,Bibliography}

\end{document}